\documentclass[11pt,showpacs,showkeys,nofootinbib,preprintnumbers]{revtex4-2}
\usepackage{graphicx} 
\usepackage{color} 
\usepackage{braket}
\usepackage{amsmath}
\usepackage{amssymb}
\usepackage{mathrsfs}
\usepackage{titlesec}
\usepackage{makecell}
\usepackage{natbib,hyperref}
\usepackage[mathscr]{euscript}
\usepackage{dutchcal}
\usepackage{mathrsfs}
\usepackage{bm}
\usepackage[T1]{fontenc}
\usepackage[utf8]{inputenc}
\usepackage[english]{babel}

\def \bea {\begin{eqnarray}}
\def \eea {\end{eqnarray}}

\newcommand{\nn}{\nonumber}

\definecolor{ao-english}{rgb}{0.0, 0.5, 0.0}
\definecolor{cadmiumblue}{rgb}{0.0, 0.42, 0.24}

\begin{document}

\preprint{ECTP-2026-08}
\preprint{WLCAPP-2026-08}
\hspace{0.05cm}

\title{Quantum-Conditioned Curvatures in Spacetime Surrounding Kerr--Newmann Black Hole}
\email{Corresponding author: atawfik@acu.edu.eg; 400778@iu.edu.sa; atawfik@bnl.gov}

\author{A. Tawfik$^{1,2}$, Saleh O. Allehabi$^{1}$, M. Ur Rehman$^{1}$, A. A. Alshehri$^{3}$}

\affiliation{$^{1}$Department of Physics, Faculty of Science, Islamic University of Madinah (IU), Madinah 42351, Saudi Arabia}

\affiliation{$^{2}$Basic Science Department, Faculty of Engineering, Ahram Canadian University (ACU), Giza 12556, Egypt}

\affiliation{$^{3}$Department of Science and Technology, University of Hafr Al Batin (UHB), Nairiyah 31981, Saudi Arabia}

\begin{abstract}
This research examines the possibility whether the curvatures found in conventional General Relativity (GR) are the only existing ones, using both analytical and numerical techniques. To this end, we introduce a thorough investigation of Riemann curvatures in the spacetime surrounding a Kerr-Newmann black hole, which is distinguished by its specific electric charges and rotational dynamics. We apply a geometric quantization ansatz that centers on the quantization of the metric tensor, from which the complete set of field equations can be derived. The conformal transformation of the standard metric tensor upholds all the principles of GR while also extending its applicability to lower (quantum) scales. We recognize two types of Riemann curvatures. In addition to the positive curvatures present in classical GR formulations, we also find significant negative curvatures at lower (quantum) scales. This may indicate quantum sources of gravitation that classical GR does not seem equipped to explore.
\end{abstract}


\date{\today}

\maketitle


\section{Introduction}
\label{sec:intro}

The essential proposition that the fundamental metric tensor combines essential information about the underlying geometry in General Relativity (GR) \cite{Weyl1917,blank1999hilbert,o1983semi,Howl:2018qdl,Sidharth:1997gxz,DeWitt:1967yk,Ashtekar:1986yd} was the key motivation for a new geometric quantization approach
\cite{Tawfik:2023rrm,Tawfik:2023kxq}. Various notions, such as duality-symmetry configurations, Finsler--Hamilton manifold \cite{chern2005riemann,xia2025geometry}, noncommutative algebra, and quantum geometry \cite{Rosen:1962mpv}, have been invented and imposed on GR
\cite{Tawfik:2024bdf,NasserTawfik:2024afw,Tawfik:2023onh,Tawfik:2023hdi}. The four-dimensional Riemann manifold has been generalized into the phase-space structured Finsler--Hamilton manifold \cite{chern2005riemann,xia2025geometry}, which allows for the integration of quantum-mechanical components. The resulting Finsler structure could be expanded to the auxiliary coordinates and momenta of a test particle, namely $x_0^{\alpha}$ and $p_0^{\beta}$, respectively. Utilizing the relativistic generalized uncertainty principle (RGUP) \cite{Tawfik:2024gxc}, which generalizes Quantum Mechanics (QM) to finite relativistic energies and gravitational fields, enables transforming the momenta operator $p_0^{\nu}$ to $\phi(x) p_0^{\nu}$, where the function $\phi(x)$ reads $\phi(x) =1+\beta p_0^{\rho} p_{0 \rho}$. The Finsler metric can be obtained from the Hessian of the resulting structure $F^2(x_0^{\alpha}, \phi p_0^{\beta})$. By equating the measures of line elements on Finsler and Riemann manifolds, one can derive the quantized four-dimensional metric on Riemann manifold \cite{Tawfik:2025wsl}. From its conformal transformation \cite{Tawfik:2024bdf,NasserTawfik:2024afw,Tawfik:2023onh,Tawfik:2023hdi}, one concludes that all postulates of GR are preserved. While the classical approximation of GR remains valid at large scales, the quantized GR extends the theory to low (quantum) scales.

The present script introduces analytical and numerical study of the classical and quantum-conditioned Riemann curvatures in the spacetime surrounding Kerr-Newmann black holes \cite{Bhattacharyya:2012wz}. Implications on other types of black holes have been reported in recent literature \cite{Tawfik:2025icy,Tawfik:2025rel,Tawfik:2025ldp,Tawfik:2025kae,NasserTawfik:2024afw}. A brief review of the formalism is outlined in Section \ref{sec:frmlsm}.

\section{Formalism}
\label{sec:frmlsm}

In GR, spacetime and the corresponding geometry are modeled as a four-dimensional Riemann manifold $(M, g_{\mu\nu})$, where $M$ is a smooth differentiable manifold and $g_{\mu\nu}$ is the metric tensor~\cite{Sasaki1958,Carroll2004Spacetime}. 
This framework is expanded to include a Finsler manifold $(M,F)$, which consists of a smooth manifold $M$ that is equipped with a Finsler structure $F:TM\to\mathbf{R}^+$, $(x,y) \mapsto F(x,y)$, where $TM$ is the tangent bundle. The Finsler structure $F(x,y)$ plays a crucial role and is assumed to hold the following properties: (i) regularity: $F$ is $C^\infty$ on $TM\setminus\{0\}$, (ii) positive homogeneity: $F(x, \lambda y) = \lambda F(x, y)$ for all $\lambda > 0$, and (iii) strong convexity: Hessian matrix $g_{ij}(x,y) = \frac{1}{2}\frac{\partial^2 F^2(x,y)}{\partial y^i \partial y^j}$ is positive definite \cite{chern2005riemann,xia2025geometry}. With RGUP and preserving the homogeneity property, Finsler structure is the quantity in which quantum-mechanical ingredients can be integrated. After translating Finsler into Riemann line element, the non-truncated four-dimensional metric tensor reads
\bea
\tilde{g}_{\mu \nu}(x,p)  &=& \left(\phi^2(x)+2\frac{\kappa}{(p_0^0)^2} F^2(x,p)\right) 
 \left[1 + \frac{\dot{p}_0^{\alpha} \dot{p}_0^{\beta}}{{\mathscr F}^2} \left(1+2\beta p_0 ^{\rho} p_{0 \rho}\right) \right] g_{\mu\nu} \nn \\
&+& \left[\frac{d x_0^{\alpha}}{d \zeta^{\alpha}} \frac{d x_0^{\beta}}{d \zeta^{\beta}} + \left(1+2\beta p_0 ^{\rho} p_{0 \rho}\right) \frac{d p_0^{\alpha}}{d \zeta^{\alpha}} \frac{d p_0^{\beta}}{d \zeta^{\beta}} \right] d_{\mu\nu}, \hspace*{5mm} \label{eq:gmunuQ2c}
\eea
where $\kappa=\beta/(p_0^0)^{2}$ with $\beta$ is the RGUP parameter, $\zeta^{\mu}$ represents parametrization that connect the coordinates in the Finsler tangent or Hamilton cotangent bundle to the Riemann coordinates, and ${\mathscr F}$ is the maximum proper force \citep{Tawfik:2023orl} that allows quantum particle to gain maximum proper acceleration along its course in the additional quantum curvatures \cite{Caianiello:1981jq,caianiello1984maximal,Caianiello:1989wm,Caianiello:1989pu,brandt1989maximal}. It is obvious that $\dot{p}_0^{\mu} \dot{p}_0^{\nu}$ are the squared force acting on that quantum particle. For the sake of simplicity, let us assume that $\tilde{g}_{\mu \nu}(x,p)$ can be truncated and the conformal coefficient is given as 
\bea
C(x,p)&=& \left(\phi^2(x)+2\frac{\kappa}{(p_0^0)^2} F^2(x,p)\right) \left[1 + \frac{\dot{p}_0^{\alpha} \dot{p}_0^{\beta}}{{\mathscr F}^2} \left(1+2\beta p_0 ^{\rho} p_{0 \rho}\right) \right], \label{eq:Ctildegalphabeta1}
\eea
Thus, the quantization of the metric tensor is considered as suitable approach for quantizing GR 
\bea
\tilde{g}_{\mu\nu}(x,p) &=& C(x,p)\, g_{\mu\nu}, \label{eq:tildegalphabeta1}
\eea
where the conformal coefficient $C(x,p)$ includes all components imposed by QM that are applied to the quantized metric tensor. Formulating $C(x,p)$ with precision presents a mathematical challenge so far. While attempting to resolve this challenge, the expression indicated in Eq. \eqref{eq:tildegalphabeta1} is an approximation that, for the time being, appears to be unavoidable. 

Having derived the quantized metric tensor, the Levi--Civita or affine connections can be deduced, straightforwardly,
\bea
\tilde{\Gamma}^{\mu}_{\delta \nu}= 
\Gamma^{\mu}_{\delta \nu}+\frac{F^{2}_{, \gamma}}{2 C(x,p)}\left(\delta^{\mu}_{\nu}+\delta^{\mu}_{\delta}-g^{\mu \gamma} g_{\delta \nu}\right). \label{eq:C}
\eea
For the sake of simplicity, let us assume that the simplest Finster metric, Klein metric \citep{Klein1910}, expresses the corresponding structure
\bea
F^{2}_{, \gamma} &=& 2 \frac{\left\langle x_0^{\mu} \cdot p_0^{\nu}\right\rangle}{\left[\left(x_0^{\mu}\right)^2 -1\right]^2} \left\{\left[1-\left(x_0^{\mu}\right)^2\right] p_0^{\nu} - x_0^{\mu} \left\langle x_0^{\mu} \cdot p_0^{\nu}\right\rangle \right\}. \hspace*{5mm}
\eea
Both analyses presented herein differentiate between the classical and quantum-conditioned results of the proposed geometric quantization. The main focus is on the Riemann curvatures \cite{Brandt:1991hy}, which can be expressed in terms of the Levi--Civita connections \cite{Brandt:1991hy},
\bea
\tilde{R}^{\gamma}_{\beta\mu\nu} &=& R^{\gamma}_{\beta\mu\nu} + \tilde{\Gamma}^{\gamma}_{\sigma \mu} + \tilde{\Gamma}^{\gamma}_{\sigma \nu , \mu} + \tilde{\Gamma}^{\gamma}_{\sigma \mu , \nu} -\left(\Gamma^{\gamma}_{\sigma \nu , \mu} + \Gamma^{\gamma}_{\beta \mu , \nu} \right) \nn \\
&+& \frac{F^2_{, \gamma}}{2C(x,p)} \left(g^{\alpha \gamma}_{, \mu} g_{\sigma\nu} + g^{\alpha \gamma} g_{\sigma\nu , \mu} + g^{\alpha \gamma}_{, \mu} g_{\beta\mu} + g^{\alpha \gamma} g_{\beta\mu , \nu} \right). \hspace*{5mm} \label{eq:RiemannQnt}
\eea
In this regard, the unquantized quantities reads
\bea
\Gamma^{\gamma}_{\mu\nu} &=& \frac{1}{2} g^{\beta\gamma} \left(g_{\mu\beta, \nu} + g_{\beta \nu, \mu} - g_{\mu \nu, \beta}\right), \label{eq:LCclss} \\
R^{\gamma}_{\beta\mu\nu} &=& \Gamma^{\gamma}_{\sigma \mu} + \Gamma^{\gamma}_{\sigma \nu , \mu} + \Gamma^{\gamma}_{\sigma \mu , \nu} -\left(\Gamma^{\gamma}_{\sigma \nu , \mu} + \Gamma^{\gamma}_{\beta \mu , \nu} \right). \label{eq:Riemann}
\eea

The numerical results obtained for charged, massive, rotating, spherically symmetric black hole, the Kerr--Newmann metric, are discussed in Section \ref{sec:Rslts}.

\section{Results}
\label{sec:Rslts}

Similar to the Reissner--Nordstr{\"o}m solution, which describes a massive, charged, {\it non-rotating}, spherically symmetric black hole, the static solution to the Einstein--Maxwell field equations for the gravitational field of a massive, {\it charged}, rotating, spherically symmetric black hole, known as the Kerr--Newmann metric, indicates that the line element in Boyer--Lindquist coordinates $(t,r,\theta,\phi)$ can be expressed as 
\bea
ds^2 &=& -\frac{c^2 \Delta(r) - c^2 a^2 \sin^2(\theta)}{\Sigma^2(r,\theta)} d t^2
+ \frac{\Sigma^2(r,\theta)}{\Delta(r)} d r^2 + \Sigma^2(r,\theta) d \theta^2  \nn \\ 
&+&  \left\{a^2 \Delta(r) \sin^2(\theta) -\left[r^2+a^2\right]^2\right\}\frac{\sin^2(\theta)}{\Sigma^2(r,\theta)} d \phi^2 \nn \\
&+& 2 c a \left[\left(r^2+a^2\right)-\Delta(r)\right] \frac{\sin^2(\theta)}{\Sigma^2(r,\theta)}. dr d\phi,
\eea
where $\Delta(r)=r^2 - r_S r + a^2 +r_Q^2$ and $\Sigma^2(r,\theta)=r^2+a^2 \sin^2(\theta)$. The length scale $a=J/(Mc)$, where $M$ is the mass of Kerr--Newmann black hole and $J$ is the angular momentum which characterizes its rotation. The Schwarzschild radius $r_S=2GM/c^2$, where $G$ is the gravitational constant. The length scale corresponding to the electric charge $Q$ is given as $r_Q=G Q^2/(4 \pi \epsilon_0 c^4)$, where $\epsilon_0$ is the vacuum permittivity. One might assume geometric units so that $G=c=\cdots=1$.

The Kerr--Newmann black hole is distinguished by its angular momentum and electric charge. It features distinct properties in contrast to non-rotating non-charged (Schwarzschild), non-rotating charged (Reissner--Nordstr{\"o}m), and rotating non-charged (Kerr) black holes. The rotation leads to frame-dragging and a quasispherical event horizon, in addition to an ergosphere, which is the area surrounding the black hole where spacetime is pulled. The electric charge allows for the consideration of energy associated with electromagnetic fields.

\begin{figure}[htb!] 
\includegraphics[angle=-90,width=0.9\textwidth]{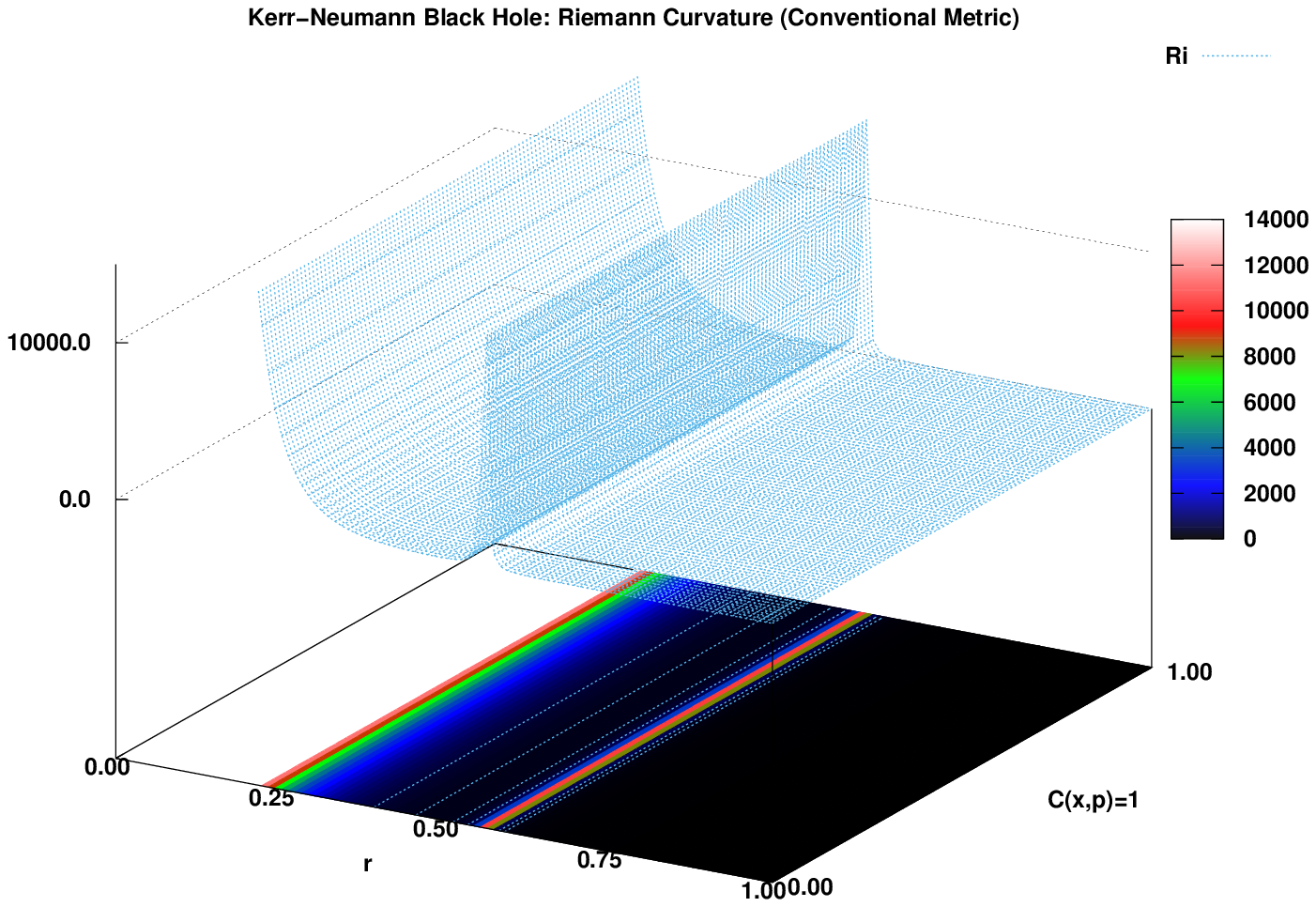}
\caption{Riemann curvatures, Eq. \eqref{eq:Riemann}, are depicted as a function of the radial distance $r$. \label{fig:1a}}
\end{figure}

The analytical expressions detailed in the preceding section are analyzed numerically, in Fig. \ref{fig:1a}. The Riemann curvatures are depicted as a function of the radial distance $r$, Eq. \eqref{eq:Riemann}. We observe that as $r$ decreases, the curvatures increase significantly, reflecting the intrinsic geometry of the underlying manifold. Riemann curvatures are characterized by the Levi-Civita connections, which is further expressed through the curvature tensor, Eq. \eqref{eq:LCclss}. The Riemann curvatures measure the noncommutativity of the covariant derivative. Positive results indicate a transition from flat to positive curvatures at all points.
 
\begin{figure}[htb!] 
\includegraphics[angle=-90,width=0.9\textwidth]{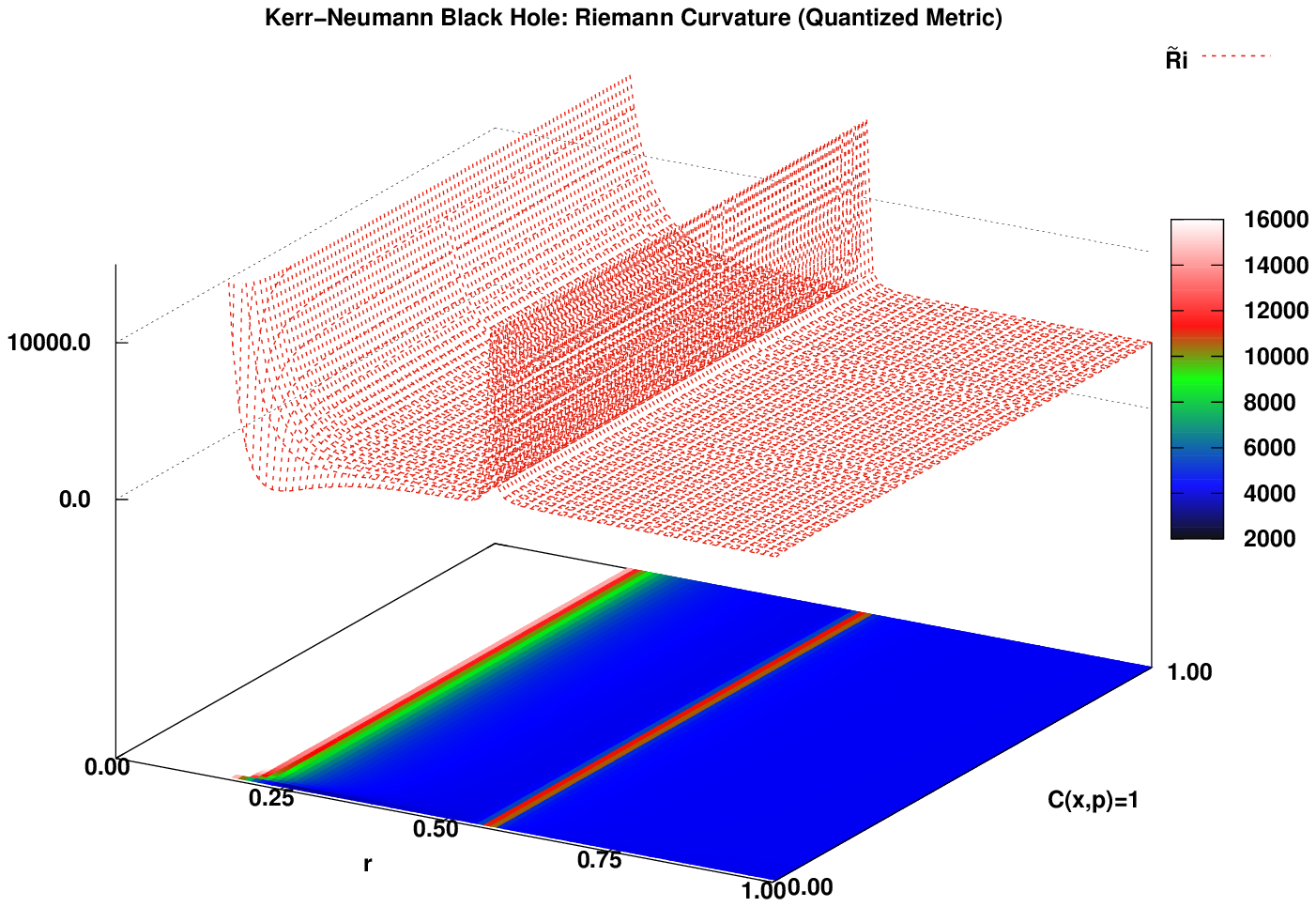}
\caption{The same as in Fig. \eqref{fig:1a} but here the results are obtained with quantized metric tensor, Eq. \eqref{eq:RiemannQnt}. \label{fig:1b}}
\end{figure}

\begin{figure}[htb!] 
\includegraphics[angle=-90,width=0.9\textwidth]{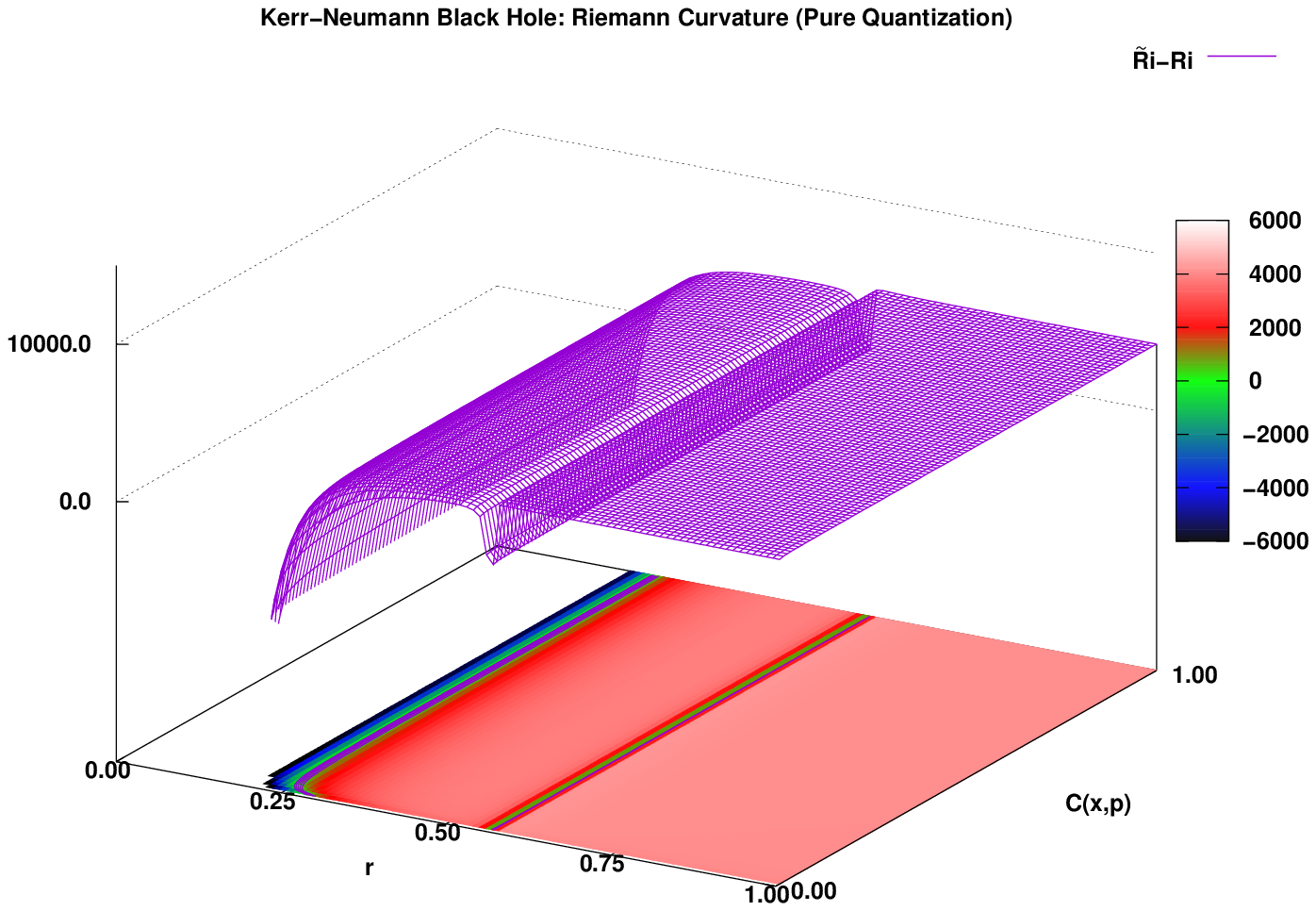}
\caption{The same as in Fig. \eqref{fig:1b} but here the curvatures exclusively stemming from the  quantization are illustrated. \label{fig:1c}}
\end{figure}

Figure \ref{fig:1b} shows the same as in Fig. \ref{fig:1a} but this instance involves a quantized metric tensor,  Eq. \eqref{eq:RiemannQnt}. As before, a decrease in $r$ is associated with a rapid rise in the Riemann curvatures. In this context, we refer to the conformal transformation presented in Eq. \eqref{eq:tildegalphabeta1} and Eq. \eqref{eq:C}, which reveals that the results discussed here linearly combine both classical and quantum-conditioned curvatures. Figure \ref{fig:1c} quantifies the contributions that are derived exclusively from the proposed geometric quantization. In this scenario, we note that lowering $r$ is linked to a rapid decline. The negative curvatures are associated with a manifold where, at a given point, the tangent space is bent away from the center of the manifold. Such negative curvatures are likely related to surfaces, for instance, that take on a saddle shape, like the hyperbolic plane, where the geodesics are represented as straight lines or semicircles that are perpendicular to the real axis. In this regard, we highlight that such negative curvatures merely arise at low (quantum) scales. This is obscured by the classical formulation of GR. The curvatures that markedly deviate from those defining classical GR indicate that a fine structure of the manifold becomes evident only at low scales. Furthermore, the proposed geometric quantization seems to facilitate an exploration of the manifold's deep structure. The finite variation that emerges before the one at very small $r$ refers to the outer ergosurface. Detailed analyses could be undertaken in other contexts.

\section{Conclusions}
\label{sec:cncl}

We have introduced an analytical and numerical assessment of the Riemann curvatures in the spacetime surrounding a Kerr--Newmann black hole. We make a distinction between classical and quantum-conditioned curvatures. This investigation was enabled by a geometric quantization ansatz that focuses on a crucial quantity in GR, specifically the metric tensor. The analyses show that the Riemann curvatures increase as the radial distance decreases, particularly in the classical formulation of GR. The proposed quantization appears to unveil fine structures and disclose another type of curvatures that are only achievable at low (quantum) scales. These curvatures significantly decrease as the radial distance decreases. The substantial negative curvatures are markedly distinct from those that arise in classical GR. They have almost the same magnitude but opposite sign. This suggests that a complex structure of the manifold is only revealed at lower scales. We conclude that the proposed geometric quantization seems to facilitate the exploration of the intricate structure of the underlying manifold.

\section*{Conflicts of Interest}

The authors declare that there are no conflicts of interest regarding the publication of this published article!

\section*{Dataset Availability}

All data generated or analyzed during this study are included in this published article. All of the material is owned by the authors.

\section*{Funding}

The authors declare that this research received no specific grants from any funding agency in the public, commercial, or not-for-profit sectors.

\section*{Data Availability}
The data supporting the findings of this study are available within the paper.

\bibliographystyle{unsrtnat}
\bibliography{ListOfReferences-KerrNeumann}


\appendix

\renewcommand{\thefigure}{A.\arabic{figure}} 
\setcounter{figure}{0} 
\setcounter{table}{0} 
\renewcommand{\thetable}{A\arabic{table}} 

\end{document}